\documentclass[journal]{IEEEtran}
\usepackage{textcomp}
\usepackage{cite}
\usepackage{amsmath,amssymb,amsfonts}
\usepackage{algorithm}
\usepackage{algorithmic}
\usepackage{graphicx}
\usepackage{multicol,multirow}
\usepackage{array}
\usepackage{pstricks}
\usepackage{comment}
\usepackage{txfonts}
\usepackage{placeins}

\newcolumntype{L}{>{\raggedright\arraybackslash}p}
\newcolumntype{C}{>{\centering\arraybackslash}m}

\usepackage{tikz}
\usepackage{circuitikz}
\newcommand{\asymcloud}[2][.5]{%
\begin{scope}[#2]
\pgftransformscale{0.12}%
\pgfpathmoveto{\pgfpoint{261 pt}{125 pt}} 
  \pgfpathcurveto{\pgfqpoint{78 pt}{382 pt}}
                 {\pgfqpoint{381 pt}{445 pt}}
                 {\pgfqpoint{412 pt}{450 pt}}
  \pgfpathcurveto{\pgfqpoint{577 pt}{547 pt}}
                 {\pgfqpoint{698 pt}{448 pt}}
                 {\pgfqpoint{695 pt}{356 pt}}
  \pgfpathcurveto{\pgfqpoint{940 pt}{162 pt}}
                {\pgfqpoint{700 pt}{197 pt}}
                 {\pgfqpoint{780 pt}{87 pt}}
  \pgfpathcurveto{\pgfqpoint{531 pt}{39 pt}}
                 {\pgfqpoint{298 pt}{51 pt}}
                 {\pgfqpoint{261 pt}{125 pt}}
\pgfusepath{fill,stroke}         
\end{scope}}    

\newcounter{sarrow}

\usepackage[final]{changes}
\definechangesauthor[name={Sameh}, color=red]{SA}
\definechangesauthor[name={Greg}, color=blue]{GR}
\usepackage[colorinlistoftodos]{todonotes}

\newcommand{\stkout}[1]{\ifmmode\text{\sout{\ensuremath{#1}}}\else\sout{#1}\fi}
\setdeletedmarkup{\stkout{#1}}
\begin{document}
%
\title{\replaced{Modelling Space-time Periodic Structures with Arbitrary Unit Cells Using Time Periodic Circuit Theory}{Generalized Space-time Periodic Circuits for Arbitrary Structures}}
%
%
%

\author{Sameh Y.~Elnaggar,~\IEEEmembership{Member,~IEEE,}
               and~Gregory N.~Milford,~\IEEEmembership{Senior Member,~IEEE}
\thanks{ S.Y.E with the Electrical and Computer Engineering Department, Royal Military College of Canada, Kingston, Ontario. G.N.M are with the School of Engineering and Information Technology, University of New South Wales, Canberra, emails: samehelnaggar@gmail.com, g.milford@adfa.edu.au}
}

\maketitle

\begin{abstract}
Using the time periodic ABCD parameters, an expression for the dispersion relation of \deleted{an arbitrary space-time modulated structure}\added{space-time modulated structures} is obtained. The relation is valid for general structures even when the spatial granularity is comparable to the operating and modulation wavelengths. In the limit of infinitesimal unit cell, the dispersion relation reduces identically to its continuous counterpart. For homogeneous space-time modulated media, the time periodic circuit approach allows the extension of the well-known telegraphist's equations. The time harmonics are coupled together to form an infinite system of coupled differential equations. At the scattering centres, where the interaction is mediated via the modulation (pump wave), the telegraphist's equations reduce to the interaction of two waves only: the signal and idler. The interaction can then be described using three wave mixing, satisfying the phase matching condition. It is demonstrated that the time periodic S parameters provide an alternative and appealing visualization of the modal conversion and the emergence of non-reciprocity inside the bandgaps.
\end{abstract}

%
\IEEEpeerreviewmaketitle

\section{Introduction}
%
%
%
%
The asymmetric interaction of space time harmonics in space-time modulated media has been recently exploited to design multitude of novel non-reciprocal devices such as magnet-less circulators \cite{AluCaloz, Alu2016magnetless, KordAlu2018}, nonreciprocal antenna \cite{Hadad2016, Taravati2017mixer,RamacciaALu2018}, one-way beam splitters \cite{TaravatiKishkarXiv}, \added{isolators via the exploitation of the asymmetric interband transition in a photonic crystal \cite{Lira2012,Winn1999} that manipulate wave transmission via a space time modulated metasurfaces \cite{Alu_PRB_2015,mazor2019}}. \added{Additionally, it was demonstrated that the inclusion of time and/or space-time periodic elements can circumvent some physical limitations of linear time invariant systems. For instance, it was shown that an appropriately time modulated reactance can result in zero reflection and hence enables extreme energy accumulation \cite{mirmoosa2019}. Quite recently, it was theoretically shown that a system based on a time switched transmission line was demonstrated to have a broadband matching capability not limited by the Bode-Fano criteria \cite{beyondbf,fano1950}. Moreover, the modulation of both the effective permittivity and permeability was shown to enhance the nonreciprocity while still keeping the spacetime modulated structure perfectly matched to the host environment \cite{Taravati2018}. For a recent review on developments, applications and methods of analysis of space-time media, please refer to \cite{taravati2019}.}

Fundamentally, the modulation of a medium constitutive parameter by a travelling wave \emph{biases} the transmission in one direction over the other(s), resulting in a skewed dispersion relation that cannot be achieved using space only or time only modulation \cite{Trainiti2016, ElnaggarMilford2018}. The theory of space-time modulated media dates back to the mid of last century when there was an immense interest in the study of distributed parametric interactions \cite{Tien1958,Cullen1958,Simon1960,Oliner1961,Cassedy1963,Cassedy1967}. During that period, the theoretical foundation was established. The framework is based on Bloch-Floquet theory, where the variables (voltages and currents or electric and magnetic fields) are represented by the infinite sum of the space time harmonics. Upon the substitution of the variables into the governing differential equation (in this case the wave equation), the system behaviour can be expressed as the interaction of the infinite number of space time harmonics. Usually only few harmonics need to be considered, particularly in the vicinity of the scattering centres \cite{ElnaggarMilford2018}. 

Practically speaking, lumped elements are used to synthesize space-time modulated structures; for instance, nonlinear elements (eg. varactors are periodically inserted in a host microstrip transmission line \cite{Qin2014}). The modulation is introduced either using a strong pump wave on the same line or via a special arrangement where the modulation comes from another coupled transmission line \cite{Qin2014,Taravati2018}. The operating regime is generally assumed and/or designed such that the granularity of the structure is small compared to the operating and modulation wavelengths; hence the structure is considered homogeneous.

Recently, it was shown that the scattering magnitude \added{in a space-time modulated Right Handed Transmission Line (RH-TL)} and \deleted{hence} \added{subsequently} the corresponding non-reciprocity can be enhanced by reducing \added{the modulation wavelength} $\lambda_m$ \cite{ElnaggarMilford2018}. \added{For a given TL length, the reduction of the $\lambda_m$ is equivalent to the increase of the interaction length.}The analysis assumed the homogeneity of the underlying structure. Additionally as $\lambda_m$ decreases so does the wavelength at the scattering centre. \deleted{For instance, at the Anti-Stokes' scattering centre the wavelength becomes $\lambda=\lambda_m/2\left(1-\nu\right)$. Additionally, the wavelength of the +1 harmonic becomes even shorter.}  Eventually, the wavelengths are reduced to the extent that they \replaced{may be}{are} just a few unit cells, raising the question about \replaced{how short the modulation wavelength can be}{the validity of the analysis and behaviour of the medium when it is becoming more like a photonic crystal}. \added{Usually, the underlying medium was implicitly assumed to be RH, where either the shunt capacitor or series inductor is modulated. }\added{The analogy between LTI and Linear Time Periodic (LTP) systems \cite{kurth1977,wereley1991} strongly suggests that an extension of the theory of periodic structures permits the exploration of arbitrary spacetime structures in a very similar fashion to that of space only periodic systems \cite{Collin2007}.} \replaced{From the previous arguments, a framework that enables the characterization of arbitrary space-time periodic structures must be based on first principles: Floquet theorem and basic circuit theory.}{To have a satisfactory answer to this and similar questions, it is imperative to take} \added{Being a circuit based approach,} the \emph{granularity} of the unit cells \added{is automatically taken} into account.\deleted{, hence the necessity of using a circuit based formalism.} \added{It is required, of course, that the} \deleted{A} circuit based approach must reduce to the homogeneous case for an infinitesimally small unit cells. It is the aim of the current manuscript to develop such framework \added{that generalizes the theory of periodic structures of LTI systems to enable the description of wave propagation in arbitrary space/time/spacetime periodic structures. For limiting cases (for instance as the modulation strength tends to zero or the structure becomes electrically small), the derived expressions reduce identically to their simpler counterparts.} \deleted{and show that it can rigorously explain the system behaviour when the wavelengths become comparable with the dimension of the unit cell and eventually tends to the homogeneous case when the unit cell becomes infinitesimally small.}

\added{The analogy between LTI and LTP circuits and systems has been exploited to analyze time periodic systems \cite{kurth1977,wereley1991,hindawi}. In RF circuits, time periodicity appears in oscillators and mixers, where a Periodic Transfer Function (PXF) is obtained after the system is linearized around the steady state limit cycle \cite{pxf}. It is worth noting that circuit based approaches have been successfully applied to describe and design time and space time periodic circuits. For instance, in \cite{estep2014,Alu2016magnetless,KordAlu2018} a circulator was built from three identical coupled resonators, where their resonant frequencies are temporally modulated and are $120^\circ$ phase shifted from one another. The resonators were realized using lumped elements, where temporal modulation is introduced via the use of varactors. In \cite{Hadad2016} a microstip line was capacitively loaded to permit coupling to freespace. The capacitances were spatiotemporally modulated to break the symmetry between absorption and emission. The dispersion relation was calculated using a generalized circuit formalism based on the cascade of time periodic cells. Hence one of the purposes of the current article is to extend such approach to arbitrary time, space and spacetime systems that are not necessarily electrically short. Such treatement permits apparently different structures to be described by the same machinery and provides a unified framework capable of characterizing the physical interactions due to the complex coupling of space-time harmonics.}

\added{The first subsection in Section \ref{sec:theory}, briefly discusses the basics of time periodic circuits to highlight the analogy with LTI systems. Using the properties of time periodic circuits, the dispersion relation of an arbtirary spacetime periodic circuit is developed in subsection \ref{subsec:stperiodic}. Additionally, under the long wavelength approximation, the space time periodic system is described by the extension of the well-known telegraphist's equations, representing space-time media by two coupled matrix equations. In Section \ref{sec:results} two examples are given presented. The first example is a RH TL, where expressions of the dispersion relation and wave behaviour of modulated homogeneous media are well understood and thoroughly described in the literature. The wave interactions of a Composite Right Left Handed transmission line (CRLH TL) are explored as a second example to demonstrate the universality of the current approach.}

\deleted{In Section \ref{sec:theory} a circuit based approach is developed for time periodic (\emph{T-periodic}) circuits, which is the direct extension of linear time invariant (LTI) circuits. Such circuits are widely used to describe nonlinear RF and microwave circuits such as mixers and oscillators, where the system is linearized around an oscillatory (limit cycle) steady state. In the context of space-time structures, such an approach was employed to calculate the dispersion characteristics of a capacitively loaded transmission line} \deleted{  Next, the spatial periodicity is included to determine the dispersion relation of an arbitrary space-time periodic structure. Under the long wavelength approximation, the \emph{T-periodic} circuit representation is shown to be the extension of the well-known telegraphist's equations, representing space-time media by two coupled matrix equations. Such equations are valid for arbitrary series and shunt lumped elements.}

\deleted{In Section \ref{sec:results}, we demonstrate the universality of the current approach by closely examining the behaviour or a right handed transmission line (RH-TL), which has been widely studied. We, however, determine the non-reciprocity when the modulation wavelength becomes just a few unit cells. It is shown that the scattering center shifts to a lower frequency due to the bending of the dispersion relation. To test the validity of the circuit method, the RH TL is solved in the time domain using a brute force Runge-Kutta method. The attenuation of the wave in the center of the bandgap is determined and compared to the theoretical prediction.}

 \deleted{Additionally, in the limit of long wavelengths (or equivalently infinitesimal unit cells), it is demonstrated that the dispersion relation tends to the one rigorously derived for homogeneous media. We employ the coupled wave formalism to the long wavelength TL to develop a system of coupled wave equations. For backward propagation, it is shown that the system of equations reduces to the well-known three wave mixing approach and is exactly equivalent to the $2\times 2$ dispersion relation. Nevertheless, the coupled wave approach provides a deeper insight into the system's internal mechanics.}
\section{Theory}\label{sec:theory}
\subsection{Theoretical Background: Time Periodic Circuits} \label{subsec:tperiodic}
\begin{figure}[!ht]
\centering
\begin{circuitikz}[scale=0.6]
\begin{scope}
 \node (cloud) at (2.45,1.68) {\tikz \asymcloud{fill=gray!20,thin,dashed,};};
\draw[red!50!black]
(0,-0.45) rectangle(5,3.2);
\draw[blue]
(2,2) to [/tikz/circuitikz/bipoles/length=0.4cm,R,l^=\scriptsize$\tilde{R}(t)$,color=blue] (3,2)
(2,1.5) to [/tikz/circuitikz/bipoles/length=0.5cm,L,l_=\scriptsize$\tilde{L}(t)$,color=blue] (3,1.5)
(1.75,1.75) to [/tikz/circuitikz/bipoles/length=0.35cm,C,l_=\scriptsize$\tilde{C}(t)$,color=blue](1.75,0.5)
(3.25,1.75) to [/tikz/circuitikz/bipoles/length=0.35cm,C,l^=\scriptsize$\tilde{C}(t)$,color=blue](3.25,0.5)
(2,2) to (2,1.5)
(3,2) to (3,1.5);
\draw[blue]
 (3,1.75) to  (3.5,1.75)
(2,1.75) to (1.5, 1.75);
\draw[blue]
(1.5,0.5) to (3.5,0.5);
\draw (-1,2.2) [short,o-,i=\scriptsize$\mathbf{I}_1$] to (0,2.2);
\draw (0,0.5) [short,-o] to (-1,0.5);
\draw (6,2.2) [short,o-,i=\scriptsize$\mathbf{I}_2$] to (5,2.2);
\draw
(5,0.5) [short,-o] to (6,0.5);
\draw (-0.7,1.35) node{\scriptsize$\mathbf{V}_1$} (5.7,1.35) node{\scriptsize$\mathbf{V}_2$};
\draw (2.4,-1.0) node{\scriptsize$\tilde{C}(t)=\tilde{C}(t+T_m),~\tilde{L}(t)=\tilde{L}(t+T_m),$} (2.5,-1.5) node{\scriptsize$\tilde{R}(t)=\tilde{R}(t+T_m)$};
\draw (-1.4,1.6)node{\Large$\rightsquigarrow$} (-1.4,1.8) node{\scriptsize$\mathbf{a}_1$};
\draw (-1.4,1) node{\Large$\leftsquigarrow$} (-1.4,0.6) node{\scriptsize$\mathbf{b}_1$};
\draw (6.2,1.6)node{\Large$\rightsquigarrow$} (6.2,1.9) node{\scriptsize$\mathbf{b}_2$};
\draw (6.2,1) node{\Large$\leftsquigarrow$} (6.2,0.65) node{\scriptsize$\mathbf{a}_2$};
\draw (2.5,-2) node{(a)};
\end{scope}
\begin{scope}[shift={(10,1.5)},scale =1.6]
\draw (0,0) circle(1 cm);
\draw [dotted] (0,-1.4)--(0,1.4);
\draw [dashed,green!50!black,->] (-1,0)--(1,0);
\draw [green!50!black](-1.4,0.1) node{$\Large\rightsquigarrow$} (-1.4,0.3) node{\scriptsize $a_0$};
\draw [green!50!black](-1.4,-0.1) node{$\Large\leftsquigarrow$} (-1.4,-0.4) node{\scriptsize $S_{11}^{(0,0)}a_0$};
\draw [green!50!black](1.2,0) node{$\Large\rightsquigarrow$} (1.4,0.3) node{\scriptsize $S_{21}^{(0,0)}a_0$};
\draw [blue!50!black,dashed](0,0)--(-0.5,0.866)  (-0.55,0.966) node[rotate=120]{$\Large\rightsquigarrow$} (-0.6,1.3) node{\scriptsize $S_{11}^{(1,0)}a_0$};
\draw [red!50!black,dashed](0,0)--(-0.5,-0.866)(-0.55,-0.966) node[rotate=-120]{$\Large\rightsquigarrow$}(-0.6,-1.3) node{\scriptsize $S_{11}^{(-1,0)}a_0$};
\draw [blue!50!black,dashed](0,0).. controls (0.55,0.3)..(0.708,0.708) (0.8,0.8) node[rotate=45]{$\Large\rightsquigarrow$}(1,1.3)node{\scriptsize $S_{21}^{(1,0)}a_0$};
\draw [red!50!black,dashed](0,0).. controls (0.55,-0.3)..(0.708,-0.708) (0.8,-0.8) node[rotate=-45]{$\Large\rightsquigarrow$}(1,-1.2)node{\scriptsize $S_{21}^{(-1,0)}a_0$};
\draw  [gray!70!black,->](0,1.4)--(-0.25,1.4);\draw [gray!10!black] (-0.125,1.6) node{\scriptsize Port 1};
\draw [gray!70!black,->](0,-1.4)--(0.25,-1.4);\draw[gray!10!black] (0.125,-1.6) node{\scriptsize Port 2};
\draw (0,-2.3) node{(b)};
\end{scope}
\end{circuitikz}
\caption{(a) A generic \emph{T-periodic} circuit, consisting of an arbitrary number of \emph{T-periodic} inductors, capacitors and resistors. (b) Scattering from a generic time periodic circuit represented by the circle. The vertical dotted line conceptually separates between ports 1 and 2. }
\label{fig:TCircuits}
\end{figure}
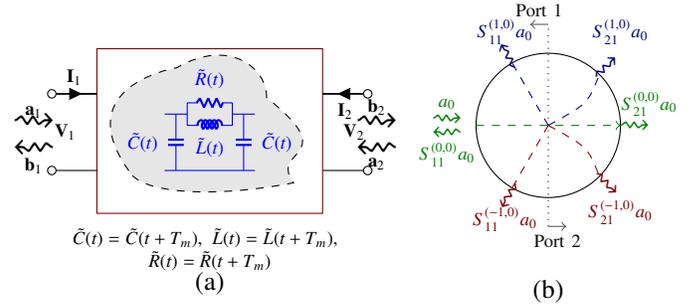
In this subsection, we \added{briefly discuss time periodic circuits. Hence the results of the subsection encompass spacetime periodic structures as well. It is worth noting that, although time periodic circuits have been investigated long time ago \cite{kurth1977}, it has recently gained immense interest due to its intriguing properties such as the ability to anhilate reflections using reactive loads\cite{mirmoosa2019} and circumventing the Bode Fano bound \cite{beyondbf}.} \deleted{dealing with circuits having elements that are \emph{T-periodic}. In this case, the periodicity in the spatial domain is not taken into account. Hence, the analysis in the current subsection is valid for any time periodic circuits, \emph{regardless} of its behaviour in the spatial domain.}  Assume the arbitrary circuit shown in Fig. \ref{fig:TCircuits}, where some of the circuit parameters (resistances, inductances or capacitances) are time periodic with a period $T_m$, where the subscript $m$ stands for modulation. Although a time periodic two port network is shown in Fig. \ref{fig:TCircuits}(a), the analysis is readily extendible to $n$ ports networks.  \added{Given a general circuit as in Fig. \ref{fig:TCircuits}(a), KVL and KCL can be applied to describe the circuit using a system of first order differential equations in the inductor currents and capacitor voltages, which define the system state vector $\mathbf{x}$.} \deleted{Generally speaking, the circuit can described in the time domain using its states (voltages across capacitors and current through inductors)} \cite{OmarCircuits}. \deleted{The number of states is denoted by $N_s$.} \added{Generally speaking the time evolution of} \deleted{the states} $\mathbf{x}$ \deleted{are coupled and} \added{is represented} \deleted{governed} by the well known state space matrix equation
\begin{equation}
\label{eq:ssm}
\dot{\mathbf{x}}=\mathbf{\mathcal{A}}(t)\mathbf{x}+\mathbf{\mathcal{B}}(t)\mathbf{u}(t).
\end{equation}
\deleted{where} \added{Due to the presence of time periodic elements in the circuit,} $\mathbf{\mathcal{A}}(t), \mathbf{\mathcal{B}}(t)$ are real time periodic matrices \deleted{that are}determined from the circuit topology via KVL/KCL and $\mathbf{u}$ is the input vector excitation. \deleted{The dimensions of $\mathbf{\mathcal{A}}$ and $\mathbf{\mathcal{B}}$ are $N_s\times N_s$ and $N_s\times 1$ (assuming a single input excitation), respectively.} Using \deleted{the} Floquet theorem, for sinusoidal excitations with frequency $\omega$, an arbitrary state $x_l$ will attain the form \cite{ElnaggarJAP2018,Miller1982}.
\begin{equation}
\label{eq:floquet}
x_l(t)=p(t)e^{i\omega t}+c.c,
\end{equation}
where $p(t)=p(t+T_m)$ is a periodic function and $c.c$ stands for the complex conjugate. Since $p(t)$ is periodic, (\ref{eq:floquet}) can written as
\begin{equation}
\label{eq:floquet2}
x_l(t)=\sum_{k=-\infty}^\infty P_ke^{i\tilde{\omega}_kt}+c.c,
\end{equation}
where $P_k$ is the amplitude of the $k^\textnormal{th}$ harmonic of $p(t)$ and $\tilde{\omega}_k\equiv \omega +k\omega_m$. According to this notation, input frequency \added{$\omega=\tilde{\omega}_0$}\deleted{$\omega=\tilde{\omega}_k$ when $k=0$}, and  hence $\omega$ and $\tilde{\omega}_0$ are synonymously used hereafter.

Substituting (\ref{eq:floquet2}) back in (\ref{eq:ssm}) and noting that $\mathbf{x}$ represents currents and voltages, the coefficients of the $p^\textnormal{th}$ harmonic can be matched, resulting in a system of \deleted{$N_s$} linear algebraic equations. The time periodic elements embedded in $\mathbf{\mathcal{A}}$ and $\mathbf{\mathcal{B}}$  couple the $p^\textnormal{th}$ harmonic with other time harmonics. In the limiting case of  linear time invariant (LTI) systems, the time harmonics are decoupled, allowing the state space variables to be determined at any given frequency $\omega$, with no knowledge of the behaviour at other frequencies. For the general time periodic case, there is an infinite number of such equations (a set for each frequency $\tilde{\omega}_p$). Practically, only a few such harmonics are necessary to characterise performance. For instance if one is interested in the behaviour at the fundamental frequency $\tilde{\omega}_0$, the first finite time harmonics $2N+1$ ($|k|\leq N$) need only be considered, where $N$ is usually small ($ 1, 2 \textnormal{ or} 3$).\deleted{ Therefore, the problem is reduced to the solution of $(2N+1)\times N_s$ algebraic equations.}

To elucidate the general approach, a simple circuit that consists of one shunt time periodic capacitance is analysed. Not only does it provide insight into the process, but also demonstrates \added{how} the concept of divide and conquer \added{is applied}, where complex systems can be divided into simpler cascaded subsystems. This concept is very useful in describing space-time periodic circuits, as will be shown in the next subsection. The current through the time periodic capacitance $\tilde{\mathcal{C}}$ is given by
\begin{equation}
\label{eq:ivCAP}
i(t)=\frac{d}{dt}\tilde{C}(t)v(t).
\end{equation}
\deleted{Since $\tilde{\mathcal{C}}(t)$ is time periodic, it can be expanded}\added{The periodicity of  $\tilde{\mathcal{C}}(t)$ allows its expansion} in a Fourier series
\begin{equation}
\label{eq:tperiodicCAP}
\tilde{C}(t)=\sum_{k=-\infty}^\infty \mathcal{C}_ke^{ik\omega_mt}.
\end{equation}
Since $\tilde{\mathcal{C}}(t)$ is real, $\mathcal{C}_q=\bar{\mathcal{C}}_{-q}$. Furthermore using (\ref{eq:floquet2}), both $v(t)$ and $i(t)$ can be written as
\begin{equation}
\label{eq:iv}
v(t)=\sum_{r=-\infty}^\infty V_re^{i\tilde{\omega}_r t}+c.c~\textnormal{and } i(t)=\sum_{r=-\infty}^\infty I_re^{i\tilde{\omega}_r t}+c.c
\end{equation}
Substituting (\ref{eq:tperiodicCAP}) and (\ref{eq:iv}) in (\ref{eq:ivCAP}), and matching the $\exp{\left(i\tilde{\omega}_rt\right)}$ \deleted{give}\added{yield}
\begin{equation}
\label{eq:iv_cap}
I_p=\sum_{l=-\infty}^{\infty}i\tilde{\omega}_p\mathcal{C}_{p-l}V_l=\sum_{l=-\infty}^\infty Y_{p-l}V_l,
\end{equation}
where $Y_{p-l}\equiv i\tilde{\omega}_p\mathcal{C}_{p-l}$ can be interpreted as the admittance connecting the $l^\textnormal{th}$ harmonic voltage with the $p^\textnormal{th}$ harmonic current. In another words, the $p^\textnormal{th}$ harmonic current is the sum of all currents due to \emph{all} harmonic voltages. Defining the time periodic current and voltage as the infinite-dimensional vectors $\mathbf{I}=\left[\cdots,I_{p-1},I_{p},I_{p+1},\cdots\right]^t$ and $\mathbf{V}=\left[\cdots,V_{p-1},V_{p},V_{p+1},\cdots\right]^t$, (\ref{eq:iv_cap}) can be compactly written in a matrix form as
\begin{equation}
\label{eq:iv_CAP}
\mathbf{I}=\mathbf{Y}\mathbf{V}=i\Omega\mathbf{\mathcal{C}}\mathbf{V},
\end{equation}
where $\mathbf{Y}$ is the time periodic admittance, $\Omega$ is a diagonal matrix storing all harmonic frequencies $\tilde{\omega}_k$ and $\mathbf{\mathcal{C}}$ is the time periodic capacitance matrix. Similar expressions for time periodic inductors and resistors can be obtained \added{\cite{kurth1977,hindawi}}.\deleted{as shown in Table 1. It is clear from the previous analysis and Table 1 that the time periodic circuit elements relations are the direct extension of the corresponding relations in LTI networks.} As time harmonics go from $-\infty$ to $\infty$, where $0$ labels the fundamental frequency, it is convenient to number the rows and columns of the matrices with reference to the fundamental component. According to this notation, the $0^\textnormal{th}$ row denotes the fundamental component.

If a time periodic capacitance is connected in shunt between input and output terminals, the structure forms a two port network, where at $\tilde{\omega}_p$ the input voltage and current ($\mathbf{V}_1$ and $\mathbf{I}_1$) are related to the output values ($\mathbf{V}_2$ and $\mathbf{I}_2$) by the ABCD parameters,
\begin{equation}
\begin{bmatrix}
\mathbf{V}_1\\\mathbf{I}_1
\end{bmatrix}
=\begin{bmatrix}
[\mathbf{A}] & [\mathbf{B}]\\
[\mathbf{C}] & [\mathbf{D}]
\end{bmatrix}
\begin{bmatrix}
\mathbf{V}_2 \\ \mathbf{I}_2
\end{bmatrix},
\end{equation}
where $[\mathbf{A}],~[\mathbf{D}]$ are the identity matrices, $[\mathbf{B}]$ is the zero matrix and $[\mathbf{C}]$ (not to be confused with the capacitance matrix $\mathcal{C}$) is the matrix where its $(m,n)$ element is given by $C_{m,n}=Y_{m-n}$. The time periodic ABCD matrix is a generalization of the well-known ABCD matrix of LTI systems. Depending on the circuit configuration, some circuit parameters can be more convenient than others. For instance, shunt (series) elements are naturally represented by the Y (Z) parameters. The conversion process between different sets of parameters parallels that of LTI systems \cite{Pozar}.

Generally, each parameter is an infinite dimensional square matrix (hence the inner brackets), but practically only few harmonic need to be considered. In the subsequent analysis, for simpler notation, the inner brackets will be dropped. For a time periodic system, the conventional S-parameters ($S_{11},~S_{21},~S_{12}~\textnormal{and } S_{22}$) are extended to four matrices. The $\mathbf{S}_{11}$ represents the reflection coefficients at port number 1, when port 2 is terminated in the reference impedance. For instance, with reference to Fig. \ref{fig:TCircuits}(b), consider an incident wave with frequency $\omega$ and complex amplitude $a_0$. $S_{11}^{(r,0)}a_0$ represents the wave bouncing back at port 1 in the $r^\textnormal{th}$ harmonic. Similarly, $S_{21}^{(r,0)}a_0$ is the wave transmitted to port 2 in the $r^\textnormal{th}$ harmonic.

\deleted{The total current and voltage at any instant satisfy KCL and KVL, respectively. Matching the $k^\textnormal{th}$ harmonic, both KCL and KVL are satisfied for each component, i.e,}
\deleted{\begin{equation}
\deleted{\sum_{i=1}^Q \mathbf{I}_i=\mathbf{0}}
\end{equation}
}
\deleted{for all current branches $Q$ entering a given node and}
\deleted{
\begin{equation}
\deleted{\sum_{i=1}^Q\mathbf{V}_i=\mathbf{0}}
\end{equation}
}
\deleted{for all voltages through a closed loop. Here $\mathbf{0}$ stands for the null vector, emphasizing the fact that each KCL and KVL are satisfied for each time harmonic $\tilde{\omega}_k$.}
\subsection{Space-time Periodic Structures}\label{subsec:stperiodic}
The analysis in subsection \ref{subsec:tperiodic} is general for time periodic systems. However sometimes, a system is also periodic in the spatial domain, forming a travelling wave modulation. This can result, for instance, from the linearization of nonlinear distributed structures \cite{Sameh_TAP_TWM} or the Distributedly Modulated Capacitance technique \cite{Qin2014}. In this subsection, we will \added{extend the time periodic}\deleted{show how the general} circuit approach discussed in the previous section\deleted{ can be extended to the case of}\added{ to } space-time periodic structures.

\begin{figure}[!h]
\centering
\begin{tikzpicture}[scale=0.6, every node/.style={scale=0.7}]
\draw [black, very thick] (0,0) rectangle (3,3);
\draw [black, very thick] (4,0) rectangle (7,3);
\draw (1.5,1.5) node{$G\left(t-x/\nu_m\right)$};
\draw (5.5,1.5) node{$G\left(t-[x+p]/\nu_m\right)$};
\draw [black, very thick, dotted] (8,0) rectangle (11,3);
\draw [black] (3,2) -- (4,2);
\draw [black] (3,1) -- (4,1);
\draw [black,dotted] (7,1) -- (8,1);
\draw [black,dotted] (7,2) -- (8,2);
\draw [black, dotted] (11,1) -- (12,1);
\draw [black, dotted] (11,2) -- (12,2);
\draw [black,dotted] (-1,1) -- (0,1);
\draw [black,dotted] (-1,2) -- (0,2);
\draw [black, densely dotted] (3.5,3.5) -- (3.5,-0.6);
\draw [black, densely dotted] (7.5,3.5) --(7.5,-0.6);
\draw (3.5,-1) node{$n^{\textnormal{th}}$};
\draw (7.5,-1) node{$(n+1)^\textnormal{th}$};
\draw [red!40!black,->](6,-0.6) -- (7.5,-0.6);
\draw [red!40!black,->](5,-0.6) -- (3.5,-0.6);
\draw (5.5,-0.5) node{$p$};
\draw [green!20!black,->](4,3.2) -- (6.5,3.2);\draw [green!20!black](6,3.4) node{$+x$};
\draw [gray!100!black](1.5,2.5)node{\Huge$\rightsquigarrow$};
\draw [gray!100!black](1.5,2.0)node{\Huge$\rightsquigarrow$};
\draw [gray!100!black](1.5,1)node{\Huge$\rightsquigarrow$};
\draw [gray!100!black](1.5,0.5)node{\Huge$\rightsquigarrow$};
\draw [gray!100!black](5.5,2.5)node{\Huge$\rightsquigarrow$};
\draw [gray!100!black](5.5,2.0)node{\Huge$\rightsquigarrow$};
\draw [gray!100!black](5.5,1)node{\Huge$\rightsquigarrow$};
\draw [gray!100!black](5.5,0.5)node{\Huge$\rightsquigarrow$};
\draw [gray!100!black](9.5,2.5)node{\Huge$\rightsquigarrow$};
\draw [gray!100!black](9.5,2.0)node{\Huge$\rightsquigarrow$};
\draw [gray!100!black](9.5,1)node{\Huge$\rightsquigarrow$};
\draw [gray!100!black](9.5,0.5)node{\Huge$\rightsquigarrow$};
\end{tikzpicture}
\caption{An arbitrary space-time periodic structure with a unit cell of $p$ units of length. The wiggly arrows emphasize the modulation in the spatial domain.}
\label{fig:STP}
\end{figure}
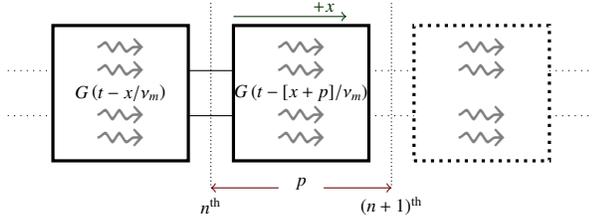

Fig. \ref{fig:STP} shows a hypothetical space-time periodic structure, where the modulation, regardless of it source, is represented by a travelling wave $G(t-x/\nu_m)$, where $\nu_m$ is the wave front velocity. \added{Accordingly, the} \deleted{The} wave front travels a unit cell length $p$ in $p/\nu_m$ units of time. The wave $G$ can be the modulated capacitance, inductance or resistance. Generally, the spatially modulated elements are placed $p$ units apart or at $x=np$, where $n$ is an integer. There can be any number of modulated elements per unit cell, for instance a shunt capacitance or both capacitance and inductance of a right handed transmission line as in \cite{Taravati2018}. In this case there will be a travelling wave $G(t-x/\nu_m)$ for each modulated element and all travel with the same speed $\nu_m$.  Any travelling wave $G$ can then be written as
\begin{equation}
G(t-x/\nu_m)=G(t'),
\end{equation}
where $t'=t-x/\nu_m$. $G(t')$ is periodic with a \deleted{minimum} period $T_m$, therefore\deleted{ it can be expanded in its Fourier components}
\begin{equation}
\begin{split}
G(t') &=\sum_{r=-\infty}^{\infty}G^r\exp(ir[\omega_mt-\beta_mx]),
\end{split}
\end{equation}
where $\beta_m\equiv\omega_m/\nu_m$ is the modulation wave number.  In the long wavelength approximation, the exact positions of the modulated elements \deleted{as well as}\added{and} the distance between them are irrelevant as long as $p$ is much smaller than the operating and modulation wavelengths. We seek a general solution \deleted{$\theta(t,x)$} that satisfies the Bloch-Floquet condition \added{(\ref{eq:floquet}) and (\ref{eq:floquet2}), but in $t'=t-x/\nu_m$. Therefore, the harmonics components of $v_{n+1}$ ( $ i_{n+1}$) \added{and} $v_n$ ($i_n$) \added{are related by}}\deleted{i.e,}
\deleted{\begin{equation}
\deleted{\theta(t,x)=\theta_0\exp(i[\omega t-\beta x])P(t'),~P(t'+T)=P(t'),}
\end{equation}}
\deleted{The periodicity of $P(t')$ allows $\theta(t,x)$ to be written as}
\deleted{\begin{equation}
\deleted{\theta(t,x)=\theta_0\exp(i[\omega t-\beta x])P(t-x/\nu_m)}\\
\deleted{=\sum_{r=-\infty}^\infty \Theta_0^r\exp\left[i\left(\tilde{\omega}_rt-\tilde{\beta}_rx\right)\right],}
\end{equation}}
\deleted{where $\Theta_0^r=\theta_0P^r$, $\tilde{\omega}_k=\omega+r\omega_m$, and $\tilde{\beta}_r\equiv\beta+r\beta_m$.
Using the above equation,}
\normalsize
\begin{equation}
\label{eq:Vn+1_Vn}
\mathbf{V}_{n+1}=\mathbf{P}_s\mathbf{V}_n,
\end{equation}
where $\mathbf{P}_s$ is diagonal with $P_s^{rr}=\exp{(-i\tilde{\beta}_rp)}$. $\mathbf{P}_s$ can be thought of as the spatial (hence the s subscript) propagator; it relates the voltage and current at $x=(n+1)p$ to those at $x=np$.

\subsubsection{Generalized Dispersion Relation}
Using  (\ref{eq:Vn+1_Vn}) the dispersion relation can be determined from the solution of

\begin{equation}
\label{eq:dispersionrelation1}
F(\omega,\beta)\equiv\left|
\begin{bmatrix}
\mathbf{A} & \mathbf{B}\\
\mathbf{C} & \mathbf{D}
\end{bmatrix}
\begin{bmatrix}
\mathbf{P}_s &\mathbf{0}\\
\mathbf{0} & \mathbf{P}_s
\end{bmatrix}-
\begin{bmatrix}
\mathbf{I} &\mathbf{0}\\
\mathbf{0} & \mathbf{I}
\end{bmatrix}
\right|=0,
\end{equation}
generally  $\omega$ and $\beta$ are complex.

\added{The dispersion relation above relies on the evaluation of an infinite determinant. In practice however, a finite set of harmonics will only be used. It is thus critical to determine the minimum number of harmonics necessary to accurately represent the harmonic interactions. As was previously shown for a right handed medium, whenever the modulation and medium speeds are sufficiently close (the sonic regime), higher harmonics with significant magnitudes always exist\cite{Oliner1961}. From a numerical procedure point of view, (\ref{eq:dispersionrelation1}) may be solved for $\omega$ or $\beta$ using an initial number of harmonics. The number of harmonics can then be increased until the relative change in $\omega$ or $\beta$ and the amplitude of the higher harmonics are below some prescribed values \cite{Cassedy1963}. It is worth noting that (\ref{eq:dispersionrelation1}) is mathematically equivalent to finding non-trivial solutions for an infinite system of homogeneous algebraic equations in infinite unknowns. A sufficient condition for an infinite determinant to converge (i.e, has a finite value) is that the sum of the off-diagonal components and the product of the diagonal elements are both absolutely convergent \cite{whittaker,mennicken,roberts1895}. Infinite determinants arise in situations intimately related to the structures studied here. In fact, it first appeared in Hill's original work in lunar motion, which resulted in periodic differential equations \cite{magnus1955,curtis2010}. This is not surprising since an infinite set of algebraic equations emerges directly whenever the Bloch-Floquet condition is imposed. For a generic system with arbitrary unit cell configuration as well as modulation properties (such as speed and wave shape), it may be challenging to determine whenever a given determinant converges. For an indepth description of systems of infinite equations in infinite unknowns please refer to \cite{riesz1913,fedorov2017,fedorov2018}.}

It is worth noting that due to the spatial periodicity, the $(k,k-s)$ element of any of the sub-matrices $\mathbf{X}^{(n)}=\mathbf{A}^{(n)},\mathbf{B}^{(n)}, \mathbf{C}^{(n)}~\textnormal{or } \mathbf{D}^{(n)}$  $n$ unit cells away is related to the $(k,k-s)$ element of the above ABCD matrix as
\begin{equation}
\label{eq:abcdkvsabcd}
X_{k,k-s}^{(n)}=e^{-is\beta_mp}X_{k,k-s}^{(0)}.
\end{equation}
Hence, given the ABCD parameters of any unit cell, the ABCD parameters of $N$ cells can be determined by cascading all unit cells
\begin{equation}
\label{eq:Cascade}
\begin{bmatrix}
\mathbf{A} & \mathbf{B}\\
\mathbf{C} & \mathbf{D}
\end{bmatrix}_\textnormal{tot}=
\prod_{k=0}^{N-1}
\begin{bmatrix}
\mathbf{A}^{(k)} &\mathbf{B}^{(k)}\\
\mathbf{C}^{(k)} &\mathbf{D}^{(k)}
\end{bmatrix}
\end{equation}
\subsubsection{Coupled Wave Equations}
\begin{figure}[!h]
\centering
\begin{circuitikz}[american voltages, scale=0.55]
\ctikzset {bipoles/length=0.6cm}
\draw 
[fill=orange!40!white](-6.5,-0.4)rectangle ++(2,0.8);
 \draw (-4.5,0) to [short](-3.5,0)
 (-3.5,0) to [short] (-3.5,-0.5) 
[fill=orange!40!white](-4,-0.5) rectangle ++(1,-2)
(-3.5,-2.5) to [short] (-3.5,-3);
\draw(-7,0) to [short,o-, i>^=$i_n$](-6.5,0);
\draw [short, i>^=$i_{n+1}$] (-3.5,0) to (-3,0);
\draw [-o](-3,0) to (-2.5,0); 
\draw[-o] (-3.5,-3) to (-2.5,-3);
\draw (-6.5,-3) to (-3.5,-3);
\node (1) at (-7.2,0) {};
\node (2) at (-7.2,-3) {};
\node (3) at (-2.7,0) {};
\node(4) at (-2.7,-3){};
\draw [->] (2) to (1);
\draw[->] (4) to (3);
\draw (-5.5,0) node{$Z(t,x)$};
\draw (-5,-1.5) node{$Y(t,x)$};
\draw
 (-7.5,0)node {$v_n$}
(-2,0) node{$v_{n+1}$}
(-4.5,-4) node{(a)};

\def \x{6.5}
\draw 
[fill=orange!40!white](-6.5+\x,-0.4)rectangle ++(2,0.8);
 \draw (-4.5+\x,0) to [short](-3.5+\x,0)
 (-3.5+\x,0) to [short] (-3.5+\x,-0.5) 
[fill=orange!40!white](-4+\x,-0.5) rectangle ++(1,-2)
(-3.5+\x,-2.5) to [short] (-3.5+\x,-3);
\draw(-7+\x,0) to [short,o-, i>^=$i_n$](-6.5+\x,0);
\draw [short, i>^=$i_{n+1}$] (-3.5+\x,0) to (-3+\x,0);
\draw [-o](-3+\x,0) to (-2.5+\x,0); 
\draw[-o] (-3.5+\x,-3) to (-2.5+\x,-3);
\draw (-6.5+\x,-3) to (-3.5+\x,-3);
\node (1) at (-7.2+\x,0) {};
\node (2) at (-7.2+\x,-3) {};
\node (3) at (-2.7+\x,0) {};
\node(4) at (-2.7+\x, -3){};
\draw [->] (2) to (1);
\draw[->] (4) to (3);
\draw (-5.5+\x,0) node{$Z$};
\draw (-5+\x,-1.5) node{$Y(t,x)$};
\draw
 (-7.5+\x,0)node {$v_n$}
(-2+\x,0) node{$v_{n+1}$}
(-4.5+\x,-4) node{(b)};

\draw 
(-7.5+0.5,-4.5-0.5) to [L,l_=$L$,i>^=$i_n$] (-4.5+0.5,-4.5-0.5)
(-4.5+0.5,-4.5-0.5) to [C,l_=$\tilde{C}(t)$] (-4.5+0.5,-7.5-0.5);
\draw[o-](-7.7+0.5,-4.5-0.5) to (-7.5+0.5,-4.5-0.5);
\draw [short, i>^=$i_{n+1}$] (-4.5+0.5,-4.5-0.5) to (-4+0.5,-4.5-0.5);
\draw [-o](-4+0.5,-4.5-0.5) to (-3.5+0.5,-4.5-0.5);
\draw[o-] (-7.7+0.5,-7.5-0.5) to (-7.5+0.5,-7.5-0.5) ;
\draw[-o] (-4.5+0.5,-7.5-0.5) to (-3.5+0.5,-7.5-0.5);
\draw (-7.5+0.5,-7.5-0.5) to (-4.5+0.5,-7.5-0.5);
\node (1) at (-7.7+0.5,-4.5-0.5) {};
\node (2) at (-7.7+0.5,-7.5-0.5) {};
\node (3) at (-3.7+0.5,-4.5-0.5) {};
\node(4) at (-3.7+0.5,-7.5-0.5){};
\draw [->] (2) to (1);
\draw[->] (4) to (3);
\draw
 (-8+0.5,-4.5-0.5)node {$v_n$}
(-2,-4.5-0.5) node{$v_{n+1}$}
(-4.5,-8.5-0.0) node{(c)};

\draw 
(-7.5+0.5+\x,-4.5-0.5) to [L,l_=$L_\textnormal{R}$,i>^=$i_n$] (-6+0.5+\x,-4.5-0.5) to [C,l_=$C_\textnormal{L}$] (-5+0.5+\x,-4.5-0.5)
(-5+0.5+\x,-4.5-0.5) to [C,l_=$\tilde{C}_\textnormal{R}(t)$] (-5+0.5+\x,-7.5-0.5)  (-4.5+0.5+\x,-4.5-0.5) to [L,l^=$L_\textnormal{L}$] (-4.5+0.5+\x,-7.5-0.5)  (-5+0.5+\x,-7.5-0.5) to  (-4.5+0.5+\x,-7.5-0.5) (-4.5+0.5+\x,-7.5-0.5)  (-5+0.5+\x,-4.5-0.5) to  (-4.5+0.5+\x,-4.5-0.5) ;
\draw[o-](-7.7+0.5+\x,-4.5-0.5) to (-7.5+0.5+\x,-4.5-0.5);
\draw [short, i>^=$i_{n+1}$] (-4.5+0.5+\x,-4.5-0.5) to (-4+0.5+\x,-4.5-0.5);
\draw [-o](-4+0.5+\x,-4.5-0.5) to (-3.5+0.5+\x,-4.5-0.5);
\draw[o-] (-7.7+0.5+\x,-7.5-0.5) to (-7.5+0.5+\x,-7.5-0.5) ;
\draw[-o] (-4.5+0.5+\x,-7.5-0.5) to (-3.5+0.5+\x,-7.5-0.5);
\draw (-7.5+0.5+\x,-7.5-0.5) to (-4.5+0.5+\x,-7.5-0.5);
\node (1) at (-7.7+0.5+\x,-4.5-0.5) {};
\node (2) at (-7.7+0.5+\x,-7.5-0.5) {};
\node (3) at (-3.4+0.5+\x,-4.5-0.5) {};
\node(4) at (-3.4+0.5+\x,-7.5-0.5){};
\draw [->] (2) to (1);
\draw[->] (4) to (3);
\draw
 (-8+0.5+\x,-4.5-0.5)node {$v_n$}
(-2+\x,-4.5-0.5) node{$v_{n+1}$}
(-4.5+\x,-8.5-0.0) node{(d)};

\draw [green!20!black,->](-3.5,-9.0) -- (1,-9.0);\draw [green!20!black](1,-9.5) node{$+x$};
\end{circuitikz}
\caption{One unit cell of  (a) a generic time periodic TL. (b) a time periodic TL, where the periodicity is included in the shunt element. (c) a RH TL with a modulated $\tilde{\mathcal{C}}$. (d) A modulated CRLH TL.}
\label{fig:RHTLuc}
\end{figure}
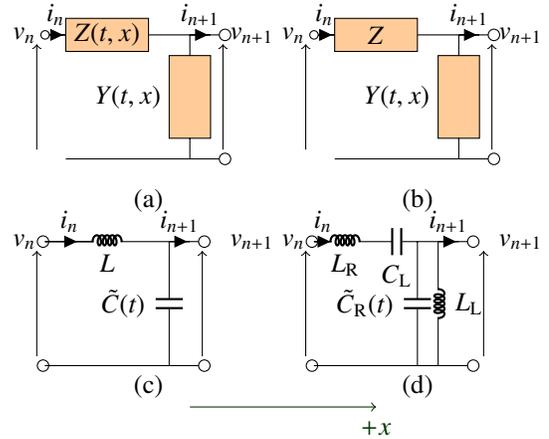

Fig. \ref{fig:RHTLuc}(a) shows a circuit that consists of repeating unit cells of arbitrary series impedance and shunt admittance. Both elements can be time modulated. This structure covers many interesting TL topologies \cite{CalozItoh, EleftheriadesBalmain}.

In the frequency domain the circuit elements are represented by the time periodic $\mathbf{Z}$ and $\mathbf{Y}$ matrices, as was shown in subsection \ref{subsec:tperiodic}. If the length of the unit cell $p$ is small compared to both the operating and modulation wavelengths $\lambda$ and $\lambda_m$ respectively, the system can be described by per unit length quantities, where generally, $\mathbf{Z}(x)=\mathbf{Z}'(x)p$ and $\mathbf{Y}(x)=\mathbf{Y}'(x)p$. Applying KVL and KCL, and in the limit of infinitesimal $p$, the voltage and current can be described using the telegraphist equations
\begin{equation}
\label{eq:coupledwave1}
\frac{d\mathbf{V}}{dx}=-\mathbf{Z}'(x)\mathbf{I} \textnormal{  and  } \frac{d\mathbf{I}}{dx}=-\mathbf{Y}'(x)\mathbf{V}.
\end{equation}

Equations (\ref{eq:coupledwave1}) are general for any time periodic structures, where $\mathbf{Z}'(x)$ and $\mathbf{Y}'(x)$ can be an arbitrary functions of $x$. This means that (\ref{eq:coupledwave1}) can be applied to spacetime periodic systems as well as to time periodic systems with an arbitrary spatial variation. By matching the boundary conditions, (\ref{eq:coupledwave1}) can be used to determine the amplitudes and phases of a given wave and its time harmonics as they propagate through a cascade of different media. It is worth noting that the state variables $\mathbf{V}$ and $\mathbf{I}$ are the time harmonics at a given position $x$. Eqs. (\ref{eq:coupledwave1}) represent an infinite system of coupled differential equations\added{, where coupling arises from the time periodic modulation}.\deleted{ The time harmonics are generally coupled due to the off-diagonal terms in $\mathbf{Z}'$ and/or $\mathbf{Y}'$, arising from the time periodic modulation.}
\section{Results and Discussion}\label{sec:results}
\deleted{The previous section proposes an expression that can be used to calculate the dispersion relation of a generic space-time modulated structure. It is the extension of the framework widely used to describe periodic structures} \deleted{, hence is readily applicable to a wide range of modulation wavelengths and frequencies.} In \deleted{this section} Subsection A, we will use the time periodic circuit approach to derive the characteristic equation of a synthesized right handed transmission line (RH TL), where the length of the unit cell can be comparable to the modulation wavelength. \deleted{Based on the derived expressions, it will be shown that when $\beta_mp\ll 1$ and for monotone modulation, the well-known dispersion relation of a homogeneous TL is automatically produced.}\added{Subsection B shed some light on the dispersion relation of a general TL, where the modulation is applied to the shunt branch. Additionally, a CRLH will be closely examined.}

\subsection{RH TL Dispersion Relation}
Similar to LTI systems, a single unit cell, shown in Fig. \ref{fig:RHTLuc} (b), can be considered as the cascade of two networks: the series LTI inductance and the shunt LTP capacitance. Hence
\begin{equation}
\label{eq:ABCD_TL}
\begin{bmatrix}
\mathbf{A}&\mathbf{B}\\
\mathbf{C}&\mathbf{D}
\end{bmatrix}=
\begin{bmatrix}
\mathbf{I}&\mathbf{Z}\\
\mathbf{0}&\mathbf{I}
\end{bmatrix}
\begin{bmatrix}
\mathbf{I}&\mathbf{0}\\
\mathbf{Y}&\mathbf{I}
\end{bmatrix}=
\begin{bmatrix}
\mathbf{I}+\mathbf{ZY} &\mathbf{Z}\\
\mathbf{Y} &\mathbf{I}
\end{bmatrix}
\end{equation}

\begin{figure}[!ht]
\centering
\includegraphics[width=3in]{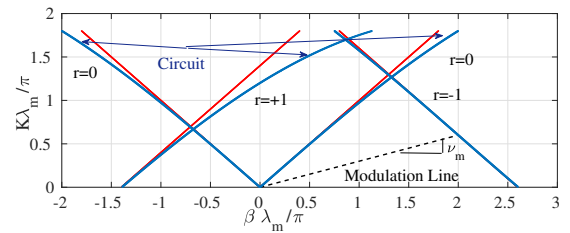}\vspace{-2mm}
\caption{Dispersion Relation for $\nu=0.3,~M\rightarrow 0$, using the time periodic circuit and long wavelength (homogeneous) approximation.}
\label{fig:Dispersion_M0p0}
\end{figure}

The matrices $\mathbf{Z}$ and $\mathbf{Y}$ are $2N+1\times 2N+1$, where the $-N$ to $N$ harmonics are only considered. The situation where the temporal periodicity is applied to one element only (usually the shunt capacitance) is very well understood and hence will be used as our test-case in the following discussion. In this case, the impedance matrix is diagonal,
\begin{equation}
\label{eq:Zkk}
Z_{kk}=i\tilde{\omega}_kL.
\end{equation}
Time periodicity appears in the $\mathbf{Y}$ matrix, where
\begin{equation}
\label{eq:Ykm}
Y_{k,m}=i\tilde{\omega}_kC_{k-m}.
\end{equation}
For monotone modulation $\tilde{\mathcal{C}}=\mathcal{C}_0\left(1+M\cos{(\omega t-\beta_mnp)}\right)$. Therefore, $\mathcal{C}_{k,k}=\mathcal{C}_0$, $\mathcal{C}_{k,k\pm 1}=M\mathcal{C}_0/2$ and $\mathcal{C}_{k,s}=0$ \deleted{$|k-s|>1$}\added{otherwise}. Substituting (\ref{eq:Zkk}) and (\ref{eq:Ykm}) in (\ref{eq:ABCD_TL}), and using (\ref{eq:dispersionrelation1}), one arrives at the system of homogeneous equations
\begin{equation}
\label{eq:threeterm}
e^{i\beta_mp}V_{k+1}+e^{-i\beta_mp}V_{k-1}+\frac{2}{M}\left[1-\left(\frac{2\sin\tilde{\beta}_kp/2}{\tilde{\omega_k}\sqrt{L\mathcal{C}_0}}\right)^2\right]V_k=0,
\end{equation}
where $k=0,\pm 1, \pm 2,\cdots$.
The system of equations (\ref{eq:threeterm}) is valid for an arbitrary modulation wavelength. The dispersion relation is determined from the non-trivial solution of (\ref{eq:threeterm}). When the modulation wavelength is considerably larger than the unit cell, i.e, $\beta_mp\ll 1$ and operating in the range where the structure is \deleted{considered} homogeneous (i.e, $\tilde{\beta}_k p\ll 1$), $2\sin\tilde{\beta}_kp/2\approx\tilde{\beta}_kp$ and (\ref{eq:threeterm}) reduces to
\begin{equation}
\label{eq:threeterm_long}
V_{k-1}+V_{k+1}+\frac{2}{M}\left[1-\left(\frac{\beta+k\beta_m}{\omega+k\omega_m}c\right)^2\right]V_k=0,
\end{equation}
where $c \equiv \lim_{p\to 0} p/\sqrt{L\mathcal{C}_0}$ is the speed of the homogeneous TL. It is convenient to normalize the wave-numbers and frequencies to be multiples of the modulation wavelength $\lambda_m\equiv 2\pi/\beta_m$. Therefore, the above equation reduces to
\begin{equation}
\label{eq:threeterm_long_normalized}
V_{k-1}+V_{k+1}+\frac{2}{M}\left[1-\left(\frac{\beta\lambda_m+2\pi k}{K\lambda_m+2\pi\nu k}\right)^2\right]V_k=0,
\end{equation}
where $K\equiv \omega/c$, the wave number of the unmodulated TL, $\nu\equiv \nu_m/c$. Equation (\ref{eq:threeterm_long_normalized}) is identical to the one derived for a modulated homogeneous medium. The circuit approach shows that such relation is only valid under the long wavelength approximations:
\begin{equation}
\beta_mp\ll 1,~ \tilde{\beta}_k p\ll 1.
\end{equation}
When the above two conditions are not satisfied, one should resort to (\ref{eq:threeterm}). This may be necessary in practical scenarios, where the TL is synthesized from \emph{finite} unit cells. Fig. \ref{fig:Dispersion_M0p0} shows the dispersion relations calculated using (\ref{eq:threeterm}) and compared to (\ref{eq:threeterm_long}) in the limit $M\rightarrow 0$. In this case, the dispersion relation of a general TL, determined by (\ref{eq:threeterm}), reduces to
\begin{equation}
2\sin\tilde{\beta}_kp/2=\pm p\tilde{\omega}_k/c 
\end{equation}
Additionally, the dispersion relation for a homogeneous TL is calculated from (\ref{eq:threeterm_long}) as
\begin{equation}
\label{eq:posAntiStokes}
\tilde{\beta}_k=\pm\tilde{\omega}_k/c 
\end{equation}
As illustrated in Fig. \ref{fig:Dispersion_M0p0}, the scattering centers are shifted down in frequency due to the bending of the dispersion curves, where $2\sin\tilde{\beta}_kp/2$ deviates from $\tilde{\beta}_kp$. \added{Bending is associated with the reduction of the group velocity, which is equal to the energy flow velocity \cite{energyflow2000}. In turn, the energy flow velocity is directly proportional to the Bloch impedance. As frequency increases, the Bloch impedance decreases (can be attributed to an increase in the equivalent capacitance of the TL), resulting in a reduction in energy flow. }The bending of the dispersion curves is particularly observed when $\beta_mp$ becomes large and is exhibited more in higher harmonics $\tilde{\beta}_k=\beta+k\beta_m,~k\geq 1$. The scattering frequencies represent the intersection of two curves given by the above equation. For instance, the first Anti-Stokes' \added{(backward)} center \cite{ElnaggarMilford2018} is determined from the intersection of the negative branch of the $0^\textnormal{th}$ curve with the positive branch of the $+1$ curve, which are given by
\begin{equation}
\label{eq:antistokeposition}
\sin\left(\beta+2\pi\right)p/2+\sin\beta p/2=\pi\nu p,~\textnormal{and } Kp=-2\sin\beta p/2.
\end{equation}

\begin{figure*}[!htb]
\centering
\includegraphics[width=5.5in]{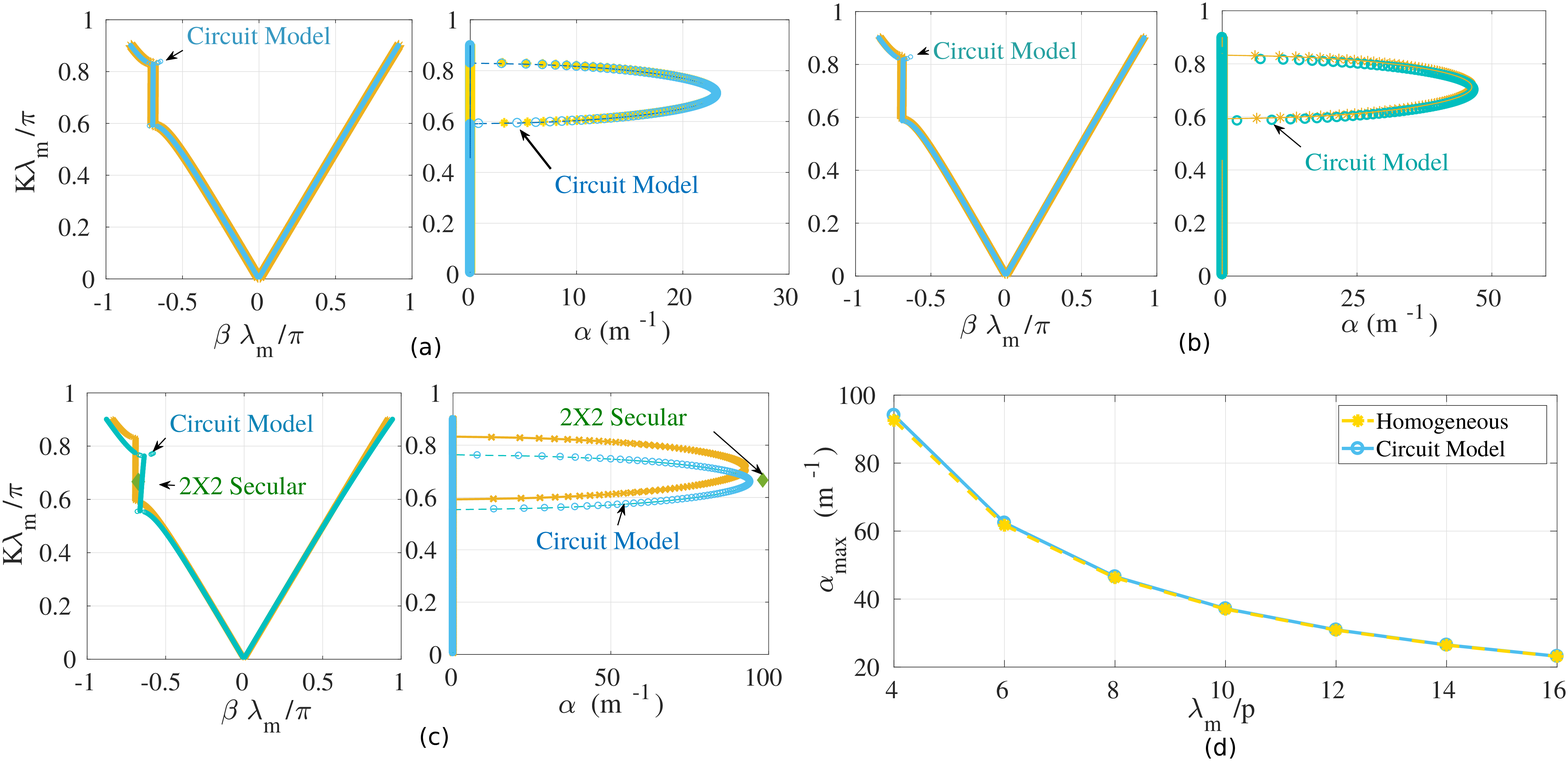}
\caption{Dispersion Relation for $\nu=0.3,~M=0.5$, using the time periodic circuit and long wavelength (homogeneous) approximation. (a)$\lambda_m/p=16$, (b) $\lambda_m/p=8$, (c) $\lambda_m/p=4$, (d) maximum $\alpha$. }
\label{fig:DispersionM0p5}
\end{figure*}
To examine the behaviour for macroscopic unit cells, the dispersion relations (\ref{eq:threeterm}) and (\ref{eq:threeterm_long}) are solved for different $\lambda_m$ values when $\nu=0.3$ and $M=0.5$. The results are reported in Fig. \ref{fig:DispersionM0p5}, where the Anti-Stokes' center is only considered. Inside the bandgap, the incident wave number become complex: $\beta-i\alpha$, where $\alpha$ is the attenuation constant. As $\lambda_m$ decreases the scattering center is shifted to a lower frequency value, as expected from the bending of the dispersion curves (Fig. \ref{fig:Dispersion_M0p0}). Since the scattering center is the intersection of the $0^\textnormal{th}$ and $1^\textnormal{st}$ harmonics, the system of equations (\ref{eq:threeterm}) can be reduced to the $2\times 2$ secular equation 
\begin{equation}
\label{eq:strengthAntiStokes}
\begin{vmatrix}
D_0 &e^{i\beta_mp}\\
e^{-i\beta_mp}&D_{+1}
\end{vmatrix}=0\textnormal{, or } D_0D_{+1}=1.
\end{equation}
The position and strength of the Anti-Stokes' center are calculated using (\ref{eq:antistokeposition}) and (\ref{eq:strengthAntiStokes}) as shown in Fig. \ref{fig:DispersionM0p5} (c). The $2\times 2$ system accurately predicts the frequency of the scattering center. However, it \emph{slightly} over estimates the value of the attenuation constant. \added{It is worth noting that, \emph{formally}, (\ref{eq:strengthAntiStokes}) has been employed to study wave propagation and radiation from modulated surfaces \cite{Oliner1959,collinzucker}.}

Additionally, Fig. \ref{fig:DispersionM0p5} (d) shows that even when $\lambda_m$ becomes just a few unit cells long, $\alpha$ is still inversely proportional to $\lambda_m$. This implies that, given a fixed modulation speed $\nu_m$, the insertion loss is directly proportional to the modulation frequency and such a relation is valid even when $\lambda_m$ is just few times larger than $p$. It is worth noting too that the behaviour for macroscopic unit cells\deleted{is}solely \deleted{dependent}\added{depends} on the bend of the dispersion curves. The coefficients $\exp{\left(-i\beta_mp\right)}$ and $\exp{\left(i\beta_mp\right)}$ appearing in the off-diagonal terms in the $2\times 2$ determinant (\ref{eq:strengthAntiStokes}) have no influence. They only affect the phase difference between the $0^\textnormal{th}$ and $+1^\textnormal{st}$  harmonics.

\begin{figure}[!htb]
\centering
\includegraphics[width=3.0in]{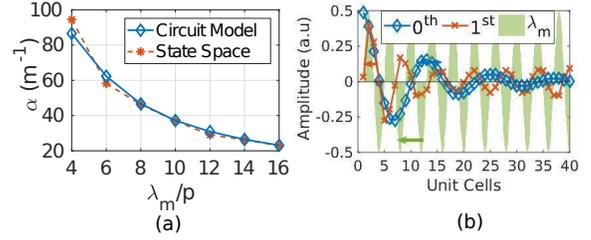}
\caption{(a) Attenuation Constant $\alpha$ calculated at the normalized frequency $Ka=\pi(1-\nu)$ using the dispersion relation and state space model. (b) Amplitude of incident and scattered signals near the Anti-Stoke's scattering center calculated using SSM. The arrows indicate the direction of propagation of the modulation and the $+1$ harmonic.}
\label{fig:Waveform}
\end{figure}

To verify the behaviour for small  $\lambda_m$ values, a time domain analysis of the transmission line is performed using a state space model (SSM) \cite{Sameh_JAP_NLD}. \added{In this case, a finite number of unit cells is used, KCL and KVL along with the circuital relations are applied in the time domain, resulting in the system of ordinary differential equations (ODEs) (\ref{eq:ssm}). The ODEs are solved using Runge-Kutta method, hence SSM enables the solution of arbitrary circuits in the time domain (regarded in this aspect as a transient solver) \cite{OmarCircuits}. From the computed time domain voltages, the magnitude and phase of the time harmonics at each node are calculated using a fast Fourier transform.} Fig. \ref{fig:Waveform} (a) shows $\alpha$ calculated using SSM and secular equation (\ref{eq:dispersionrelation1}) at the  Anti-Stokes' center ($Ka=\pi(1-\nu)$) for different values of $\lambda_m/p$. As $\lambda_m/p$ decreases, $\alpha$ increases. Fig. \ref{fig:Waveform} (b) shows the amplitude of the incident and scattered waves superimposed on the modulation wave. The modulating wavelength is four unit cells. The incident wave attenuates as it propagates through the structure due to the scattering in the +1 mode that bounces back toward the source. Furthermore when $\lambda_m/p=4$, the normalized frequency at which the maximum attenuation occurs is shifted from $0.7\pi$ to $0.66\pi$ due to the bend of the dispersion curves (Fig. \ref{fig:Dispersion_M0p0}). At this frequency the SSM predicts $\alpha_\textnormal{max}$ to be 99.57 $\textnormal{ m}^{-1}$, very close to the value predicted by the $2\times 2$ secular equation ($\alpha_\textnormal{max}=98.12\textnormal{ m}^{-1}$).

\subsubsection{Time periodic S parameters}
\begin{figure}[!htb]
\centering
\includegraphics[width=3.5in]{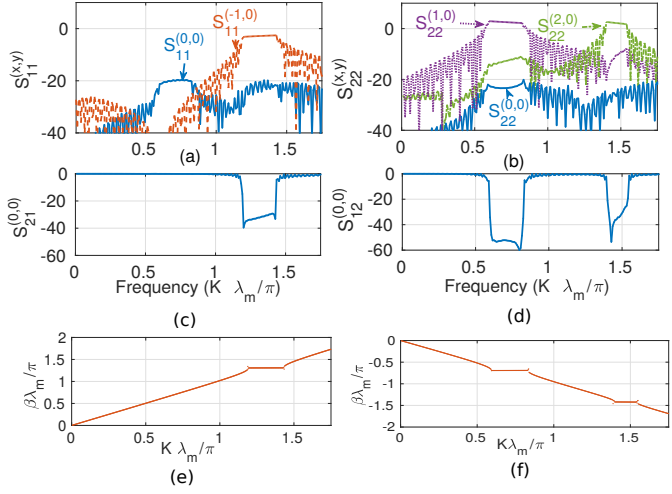}
\caption{Generalized S parameters for $\nu=0.3,~M=0.5$, using the time periodic circuit when $\lambda_m/p=8$. Forward direction: (a) scattering. (c) transmission. (e) dispersion. Backward: (d) scattering. (d) transmission. (f) dispersion.}
\label{fig:sparam1}
\end{figure}

Instead of the dispersion relation, the circuit model presents another complementary view of the interaction process that may lead to non-reciprocity. Given the number of stages\deleted{$N$}, the total ABCD matrix can be calculated using (\ref{eq:Cascade}), from which the scattering parameters are obtained. It is worth noting that the spatial modulation is only used to modify the appropriate terms in the ABCD matrices as given by (\ref{eq:abcdkvsabcd}). The time periodic S parameters are shown in Fig. \ref{fig:sparam1}. Noting that $S_{21}^{(0,0)}$ ($S_{12}^{(0,0)}$) represents \deleted{that} the scattering in the forward (backward) direction, it is clear that the isolation occurs in the bandgaps, where the transmission coefficients are significantly reduced. The $S_{11}^{(r,0)}$ parameters show that, inside the forward bandgap, scattering occurs at the $-1$ harmonic as expected ($1^\textnormal{st}$ Stokes' centers) \cite{ElnaggarMilford2018,Cassedy1963}. Additionally, scattering in the $+1$ harmonic occurs for the backward interaction ($1^\textnormal{st}$ Anti-Stokes' center). Moreover, there is another interaction at the $2^\textnormal{nd}$ backward scattering center due to the coupling between the fundamental signal and its second harmonic.

Similar to LTI systems, the scattering parameters can be used to estimate the dispersion relation. \added{Ignoring the inhomogenity  in the Bloch impedance due to the finite length of the TL and scattering in time harmonics, the forward (backward) transmission coefficient $S_{21}^{(0,0)}$ ($S_{12}^{(0,0)}$) assumes the form $S_{21}^{(0,0)}=\exp{\left([-\alpha^F-i\beta^F]L\right)}$ ($S_{12}^{(0,0)}=\exp{\left([-\alpha^B-i\beta^B]L\right)}$, where $L$ is the TL total length. Therefore knowing $S_{21}^{(0,0)}$ ($S_{12}^{(0,0)}$), enables the estimation of $\alpha^F$ and $\beta^F$ ($\alpha^B$ and $\beta^B$).} \deleted{In particular, at any given frequency, $\alpha$ and $\beta$ are estimated as}

\deleted{\begin{equation}
\label{eq:estimatedalphabetaF}
\deleted{\beta^F=\operatorname{\mathbb{I}m}\frac{\ln (S_{21}^{(0,0)})}{Np},~\alpha^F=\operatorname{\mathbb{R}e}\frac{\ln (S_{21}^{(0,0)})}{Np}}
\end{equation}
\deleted{for forward propagation and }
\deleted{\begin{equation}}
\label{eq:estimatedalphabetaB}
\deleted{\beta^B=\operatorname{\mathbb{I}m}\frac{\ln (S_{12}^{(0,0)})}{Np},~\alpha^B=\operatorname{\mathbb{R}e}\frac{\ln (S_{12}^{(0,0)})}{Np}}
\end{equation}}
\deleted{for backward propagation.} Fig. \ref{fig:Sparams} shows the \deleted{calculated}\added{estimated} $\alpha$ and $\beta$ for 100 unit cells. Qualitatively, the S parameters can be used to estimate the position  and width of the bandgaps. It is, however, not as accurate as the dispersion relations (\ref{eq:dispersionrelation1})\deleted{ mainly due to errors accumulating due to the  multiplication of the cascaded unit cells}.\added{ This is mainly due to the inhomogenity of the Bloch impedance inside the bandgap resulting from the increased scattering in time harmonics. Additionally, the finite number of unit cells used to calculate the S parameters inevitably adds to the inaccuracy of the estimation.}
\begin{figure}[!htb]
\centering
\includegraphics[width=3.0in]{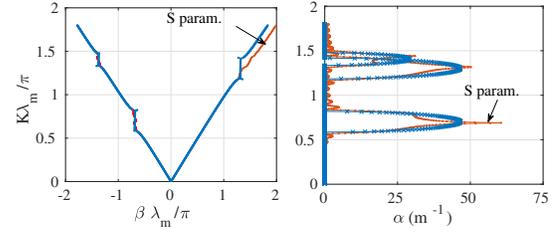}
\caption{(Left) Dispersion Relation calcuated using the secular equation (\ref{eq:dispersionrelation1}) and the time periodic S parameters. (Right) Attenation constant calculated (\ref{eq:dispersionrelation1} and S parameters. }
\label{fig:Sparams}
\end{figure}

\subsubsection{Coupled Wave Equations of spacetime periodic RH TL}
The time periodic telegraphist's equations (\ref{eq:coupledwave1}) are readily applicable to the RH TL, where the shunt capacitance is spacetime modulated. The telegraphist equations can be combined to produce the second order matrix equation
\begin{equation}
\label{eq:coupledwave3}
\frac{d^2\mathbf{V}}{dx^2}=\mathbf{Z}'(x)\mathbf{Y}'(x)\mathbf{V},
\end{equation}
which represents the interaction of an infinite number of waves. To illustrate the usefulness of the coupled wave representation (\ref{eq:coupledwave3}), the interaction of the fundamental with its $+1$ harmonic is analyzed. Such interaction is significant at the Anti-Stokes' scattering center \cite{Cassedy1963,ElnaggarMilford2018}.

For monotone modulation, using (\ref{eq:Zkk}) and (\ref{eq:Ykm}),
\begin{equation}
\label{eq:coupledEq1}
\frac{d^2V_0}{dx^2}=-\left(\frac{\omega}{c}\right)^2V_0-\left(\frac{\omega}{c}\right)^2\frac{M}{2}e^{i\beta_mx}V_1
\end{equation}
and
\begin{equation}
\label{eq:coupledEq2}
\frac{d^2V_1}{dx^2}=-\left(\frac{\omega+\omega_m}{c}\right)^2\frac{M}{2}e^{-i\beta_mx}V_0-\left(\frac{\omega+\omega_m}{c}\right)^2V_1
\end{equation}
If $V_0=A_0\exp(-i\beta x)$ it follows that $V_1=A_1\exp(-i\left[\beta+\beta_m\right]x)$; implying that the harmonics satisfy the phase matching condition. Substituting these expressions back in (\ref{eq:coupledEq1}) and (\ref{eq:coupledEq2}) results in a secular equation in $\omega$ and $\beta$
\begin{equation}
\left(\frac{2}{M}\right)^2\left[1-\left(\frac{\beta\lambda_m}{K\lambda_m}\right)^2\right]\left[1-\left(\frac{\beta\lambda_m+2\pi}{K\lambda_m+2\pi\nu}\right)^2\right]-1=0,
\end{equation}
identical to (\ref{eq:strengthAntiStokes}) under the long wavelength approximation.

\subsection{General Ladder TL}
\added{In this subsection, we demonstrate how the dispersion relation (\ref{eq:dispersionrelation1}) can be applied to a generic TL that is formed of unit cells consisting of a series and shunt impedances as shown in Fig. \ref{fig:RHTLuc}(b). Such structure covers the RH TL studies in the previous subsections and the composite right left handed (CRLH) TL presented in Fig. \ref{fig:RHTLuc}(d). The time periodicity is assumed to be purely sinusoidal and added to the shunt admittance only. Such restriction simplifies the mathematical treatement and focuses on similar behaviours of systems represented by Fig. \ref{fig:RHTLuc}(b). The sinusoidal modulation permits the reduction of the dispersion relation to the three term recursion relation \cite{meixner,lorentzen1992} }

\begin{equation}
\label{eq:general3terms}
A_kV_{k+1}+B_kV_k+C_kV_{k-1}=0,
\end{equation}
\added{where the coefficients $A_k\equiv Z_{k,k}Y_{k,k+1}\exp\left(i\beta_mp\right)$, $B_k\equiv Z_{k,k}Y_{k,k}+4\sin^2\tilde{\beta}_kp/2$ and $C_k\equiv Z_{k,k}Y_{k,k-1}\exp\left(-i\beta_mp\right)$, and the phase of the modulation is taken such that $Y_{k,k-1}=Y_{k,k+1}$. Eq. (\ref{eq:general3terms}) is the generalized form of (\ref{eq:threeterm}). Unlike (\ref{eq:threeterm_long_normalized}), first appeared in \cite{Oliner1961,Cassedy1963} for a RH media, (\ref{eq:general3terms}) is applicable to arbitrary unit cells and to situations where the unit cell length cannot be ignored. If modulation is not purely sinusoidal then (\ref{eq:general3terms}) becomes a $2N+1$ recursion, where $N$ is the number of significant modulation harmonics. It may be challenging, however, to find an analytical closed form expression for an arbitrary $2N+1$ recursion relation. Fortunately, the three term recursive relation (\ref{eq:general3terms}) can be reduced to the canonical form $y_{k+1}+y_{k-1}-D_ky_k=0$, by the change of variables: $\alpha_0=\alpha_1=1,~\alpha_{k+1}A_k=\alpha_{k-1}C_k,\forall k\geqslant 1$, $V_k=\alpha_ky_k$ and $D_k=-\left(\alpha_k/\alpha_{k+1}\right)\left(B_k/A_k\right)$ \cite{meixner}. Therefore, when }
\begin{equation}
\alpha_{2n}=\alpha_{2n+1}=e^{-i2n\beta_mp},~n=0,\pm,1,\cdots,
\end{equation}
\added{(\ref{eq:general3terms}) reduces to the canonical form, where}
\begin{equation}
D_k=-\frac{Z_{k,k}Y_{k,k}+4\sin^2\tilde{\beta}_kp/2}{Z_{k,k}Y_{k,k+1}}e^{-i\beta_mp(-1)^k}.
\end{equation}
\added{A sufficient condition that guarantees the convergence of the space-time harmonics expansion is the existence of some $N$ such that for all $n>N$, $|D_n|>2$ \cite{Oliner1959} . As was previously shown, for the three term recursion, the infinite determinant (\ref{eq:dispersionrelation1}) can be reduced to a continued fraction expansion.}

\added{As a typical example, consider the CRLH TL shown in Fig. \ref{fig:RHTLuc}(d). Here the time periodicity is introduced via the modulation of the shunt capacitance $\mathcal{C}_R$. The balanced configuration where the shunt and series resonances are equal ($\omega_{se}=1/\sqrt{L_R\mathcal{C}_L}=\omega_{sh}=1/\sqrt{L_L\mathcal{C}_R}$) is assumed to hold in the absence of the time periodicity. Since $\mathcal{C}_R$ mainly impacts the right hand regime, $Y_{k,k-1}$ is taken to be $0.8 Y_{k,k}$; hence making the interaction of space-time harmonics in the left hand regime observable. Additionally, the circuit components ($\mathcal{C}_R,\mathcal{C}_L,L_R,L_L$) are chosen such that $\omega_{se}=\omega_{sh}=1$ a.u. The modulation frequency is set to 1.2 (a.u), just above $\omega_{se}$ and $\beta_mp=0.7$. These values guarantee that the main branch ($r=0$) and the $r=\pm 1 $ branches intersect in both the left and right hand regimes (Fig. \ref{fig:crlh}(a)). The continued fraction expansion approach is used to calculate $\beta$ for any given frequency $\omega$. The continued fraction is calculated using Euler-Wallis recursive relation \cite{CFonline}. Twenty five harmonics are included to assure the convergence of the continued fraction expansion. However a much lower number of harmonics is sufficient, since the continued fraction rapidly converges.}

\added{Fig. \ref{fig:crlh}(a) and (b) depict the calculated real and imaginary parts of $\beta$, respectively. As shown, there are two main interactions in both the forward and backward directions that result in bandgaps: (1) the \emph{usual} RH-RH bandgap, appearing between $\omega=1.2-1.7$, which behaves very similar to the bandgap of a modulated RH TL. (2) The LH-LH bandgap (of a smaller magnitude appearing around $\omega=0.7$), however, is due to the interaction of the left handed wave with its $r=\pm 1$ harmonic. To reproduce the dispersion relations, a 40 unit cells CRLH TL is directly simulated in the time domain. The real and imaginary parts of the wave number are estimated from the spectrum of the time domain data. The results are presented in Fig. \ref{fig:crlh}(c) and (d). }

\added{To better explain the difference between the RH-RH and LH-LH interactions, one may refer to Fig. \ref{fig:crlh_prop}. Inside the FWD bandgaps (Fig. \ref{fig:crlh_prop}(a)), the phase velocities of the input signal and the modulation are codirectional. This means that, unlike the RH-RH interaction, the group velocity and modulation directions are contra-directional inside the LH-LH bandgap (due to the left-handensess of the medium). Therefore to observe nonreciprocity, the excitation must be applied to port 1 (P1) for a RH-RH bandgap, and to port 2 for a LH-LH bandgap(Fig.\ref{fig:crlh_prop}). Similar arguments apply to the backward situation (Fig. \ref{fig:crlh_prop}(b)).}

\begin{figure}[!htb]
\centering
\includegraphics[width=3.25in]{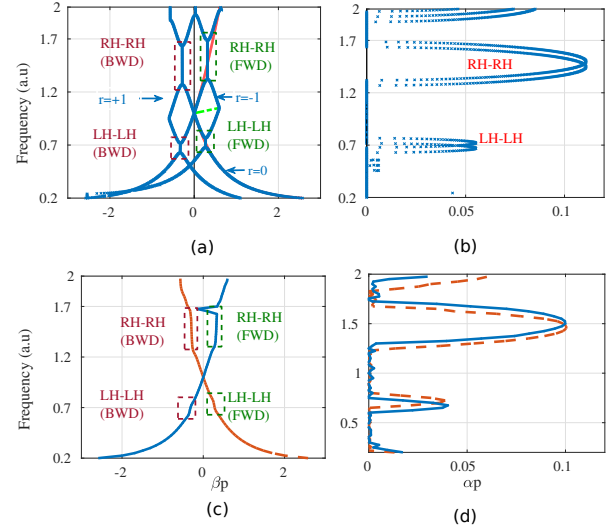}
\caption{(a), (b) Real and Imaginary of the wave number of a CRLH TL calculated using \ref{eq:general3terms}. (c), (d) Real and Imaginary of the wave number using SSM. }
\label{fig:crlh}
\end{figure}

\begin{figure}[!ht]
\centering
\begin{circuitikz}[scale=0.4]
\draw[gray!100!black]
(1,-0.45) rectangle(4,3.2);
\draw (-1.3,1.6) node{\textbf{P1}};
\draw (6.5,1.6) node{\textbf{P2}};
\draw [green!30!black](-0.2,2.3)node{\Huge$\rightsquigarrow$} (-0.4,3) node{\normalsize$a_{inc}$};
\draw [green!30!black](-0.4,3.8)node{\scriptsize RH-RH};
\draw [green!30!black](5.2,2)node{\Large$\rightsquigarrow$} (5.5,2.9) node{\normalsize$S_{21}^{(0,0)}a_{inc}$};
\draw [gray!100!black](2.5,2)node{\Huge$\rightsquigarrow$};
\draw [gray!100!black](2.5,2.7)node{\Huge$\rightsquigarrow$};
\draw [gray!100!black](2.5,1.3)node{\Huge$\rightsquigarrow$};
\draw [gray!100!black](2.5,0.6)node{\Huge$\rightsquigarrow$};
\draw [gray!100!black](2.5,-0.1)node{\Huge$\rightsquigarrow$};
\draw [green!30!black](5.2,1)node{\Huge$\leftsquigarrow$} (5.2,0.3) node{\normalsize$a_{inc}$};
\draw [green!30!black](5.2,-0.4)node{\scriptsize LH-LH};
\draw [green!30!black](-0.2,1)node{\Large$\leftsquigarrow$} (-1.0,0.4) node{\normalsize$S_{12}^{(0,0)}a_{inc}$};
\draw (2.5,-1) node{\scriptsize(a)};
\def\xshift{10}

\draw[gray!100!black]
(1+\xshift,-0.45) rectangle(4+\xshift,3.2);
\draw (-1.3+\xshift,1.6) node{\textbf{P1}};
\draw (6.5+\xshift,1.6) node{\textbf{P2}};
\draw [red!30!black](-0.2+\xshift,2.3)node{\Huge$\rightsquigarrow$} (-0.4+\xshift,3) node{\normalsize$a_{inc}$};
\draw [red!30!black](-0.4+\xshift,3.8)node{\scriptsize RH-RH};
\draw [red!30!black](5.2+\xshift,2)node{\Large$\rightsquigarrow$} (5.5+\xshift,2.9) node{\normalsize$S_{21}^{(0,0)}a_{inc}$};
\draw [gray!100!black](2.5+\xshift,2)node{\Huge$\leftsquigarrow$};
\draw [gray!100!black](2.5+\xshift,2.7)node{\Huge$\leftsquigarrow$};
\draw [gray!100!black](2.5+\xshift,1.3)node{\Huge$\leftsquigarrow$};
\draw [gray!100!black](2.5+\xshift,0.6)node{\Huge$\leftsquigarrow$};
\draw [gray!100!black](2.5+\xshift,-0.1)node{\Huge$\leftsquigarrow$};
\draw [red!30!black](5.2+\xshift,1)node{\Huge$\leftsquigarrow$} (5.2+\xshift,0.3) node{\normalsize$a_{inc}$};
\draw [red!30!black](5.2+\xshift,-0.4)node{\scriptsize LH-LH};
\draw [red!30!black](-0.2+\xshift,1)node{\Large$\leftsquigarrow$} (-1.0+\xshift,0.4) node{\normalsize$S_{12}^{(0,0)}a_{inc}$};
\draw (2.5+\xshift,-1) node{\scriptsize(b)};
\end{circuitikz}
\caption{A diagram that highlights the RH-RH and LH-LH interactions. (a) Forward Branch. (b) Backward Branch. The wiggly arrows inside the rectangle determine the direction of the modulation velocity. The attenuation of the transmitted signal is highlighted by a reduction of the wiggly arrows.}
\label{fig:crlh_prop}
\end{figure}
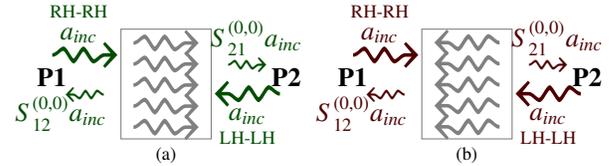

\section{Conclusion}
A circuit formalism for time periodic circuits is used to determine the dispersion relation of an arbitrary space-time periodic stucture. The relation is valid even when the length of the spatial periodicity is comparable to the modulating and operating wavelengths. In this case, the scattering centers are shifted to lower frequencies due to the bending of the dispersion relation. For infinitesimal unit cells, the relation retains the formula that was perviously derived for homogeneous media. Additionally, the system can be described by a generalized telegraphist's equations. Generally, the time harmonics are coupled by the time periodic circuit elements. The circuit based approach permits the use of S parameters to describe scattering and modal conversion in different time harmonics.

\clearpage
\bibliographystyle{IEEEtran}
\bibliography{IEEEabrv,NonReci}

\end{document}